\documentclass[aps, prd, amsmath, floats, floatfix, twocolumn, nofootinbib, superscriptaddress]{revtex4-1}
\usepackage[pdftex]{graphicx}
\usepackage{color}
\usepackage{latexsym}
\usepackage[colorlinks]{hyperref}

\newcommand{\be}{\begin{eqnarray}}
\newcommand{\ee}{\end{eqnarray}}

\begin{document}

\title{Modeling uncertainties in X-ray reflection spectroscopy measurements I:\\Impact of higher order disk images}

\author{Menglei~Zhou}
\affiliation{Center for Field Theory and Particle Physics and Department of Physics, Fudan University, 200438 Shanghai, China}

\author{Dimitry~Ayzenberg}
\affiliation{Center for Field Theory and Particle Physics and Department of Physics, Fudan University, 200438 Shanghai, China}

\author{Cosimo~Bambi}
\email[Corresponding author: ]{bambi@fudan.edu.cn}
\affiliation{Center for Field Theory and Particle Physics and Department of Physics, Fudan University, 200438 Shanghai, China}

\author{Sourabh~Nampalliwar}
\affiliation{Theoretical Astrophysics, Eberhard-Karls Universit\"at T\"ubingen, 72076 T\"ubingen, Germany}

\begin{abstract}
There are many simplifications in current relativistic reflection models and this introduces systematic uncertainties in the measurement of the properties of the source. In this paper, we study the impact of the radiation crossing the equatorial plane between the black hole and the inner edge of the accretion disk; that is, the radiation produced by the other side of the disk or circling the black hole one or more times (higher order disk images). For slow-rotating black holes with a larger plunging region, the effect is not very sensitive to the exact inclination angle of the disk. For fast-rotating black holes with a smaller plunging region, the effect is very weak for low inclination angles and becomes more important as the angle increases. We simulate some observations without and with higher order disk images and fit the data with the reflection model {\sc relxill\_nk} to check its capability of recovering the correct input parameters. Our results suggest that the effect of higher order disk images can be safely ignored for observations with present and near future X-ray missions, even for tests of the Kerr hypothesis. 
\end{abstract}

\maketitle


\section{Introduction}

Relativistically broadened reflection features are commonly observed in the X-ray spectrum of both stellar-mass and supermassive black holes~\cite{o1,o2,o3,o4,o5}. They are generated from illumination of the disk by a ``corona'', which is a generic name to indicate a hotter ($\sim 100$~keV) electron gas near the black hole, but its exact morphology is not yet well understood~\cite{rev,book}. X-ray reflection spectroscopy is potentially a powerful tool to probe the strong gravity region around black holes, measuring the black hole spin~\cite{s1,s1b,s2,s3,s4}, and potentially even testing general relativity in the strong field regime~\cite{t1,t1b,t2,t3,t4}.

Current relativistic reflection models have a number of simplifications that -- inevitably -- introduce systematic uncertainties in the final measurements of the model parameters~\cite{ss1,ss2,ss3,ss4}. When such systematic uncertainties in the theoretical model exceed the statistical uncertainties, we can get erroneous estimates of the properties of the systems. In order to get reliable measurements of the properties of black holes, it is thus crucial to have at least a rough estimate of the impact of these simplifications on the measurement of the model parameters and ensure that the corresponding systematic error is much smaller than the statistical error of the measurement.

Systematic uncertainties due to the theoretical model can be grouped into four classes: $i)$ simplifications in the calculations of the non-relativistic reflection spectrum, $ii)$ simplifications in the description of the accretion disk, $iii)$ uncertainties in the coronal properties, and $iv)$ relativistic effects not taken into account. The aim of the present paper is to investigate the impact on the reflection spectrum of a subgroup of $iv)$, which we call higher order disk images. If the plunging region between the black hole and the inner edge of the accretion disk is optically thin, a distant observer can see the radiation emitted from the other side of the disk or radiation circling the black hole one or more times as a consequence of strong light-bending. Relativistic reflection models usually ignore such radiation. Here we include this radiation and run some simulations to estimate its impact on the estimate of the model parameters when it is not included in the relativistic reflection model.

We do not expect that, among the four different classes of uncertainties, those in group $iv)$ are the most important one, nor that higher order disk images are the most important one among the class $iv)$. Nevertheless, it is necessary to have at least a rough estimate of every source of uncertainty and in the present paper we study the impact of higher order disk images. Concerning the systematic uncertainties belonging to class $iv)$, related to relativistic effects, it was argued that the effect of ``returning radiation'' may significantly affect the final measurements~\cite{andrzej1,andrzej2}. Returning radiation is radiation emitted by the disk and hitting the disk again as a result of the strong light bending near the black hole. The resulting reflection radiation is not generated by illumination of the disk by the corona, so the incident radiation is not a power-law with an exponential cut-off, but is generated by illumination of the disk by the primary reflection radiation, which is generated by the corona. Note, however, that the implementation of returning radiation into a model requires specifying the exact coronal geometry, for example a lamppost corona. The disk's intensity profile of the reflection spectrum generated by a corona with arbitrary geometry (which is the case considered in this paper) is normally modeled with a power-law or a broken power-law. In such a case, the fit automatically takes this effect into account by reabsorbing the returning radiation contribution into the emissivity index, which is the emissivity index resulting from the reflection component generated by both the corona and the returning radiation\footnote{However, this ignores the fact that the returning radiation has a reflection spectrum rather than a power-law spectrum.}.

In our study, we find that the contribution of higher order disk images is regulated by the inclination angle of the disk and by the location of the inner edge of the disk (or by the spin parameter assuming the Kerr metric and the inner edge of the disk at the innermost stable circular orbit, ISCO). For very-slow-rotating black holes (as well as for black holes with counterrotating disks), characterized by an inner edge of the disk at a relatively large radius, the contribution from higher order disk images is not very sensitive to the exact inclination angle. As the inner edge of the disk gets closer to the black hole, the contribution of higher order disk images significantly decreases for low inclination angles and slightly increases for high inclination angles. However, when the inner edge of the disk is very close to the black hole, the contribution becomes very small even for high inclination angles.

In order to be more quantitative, we simulate some observations of the reflection spectrum of an accreting black hole without and with higher order disk images. We then fit the faked data with the relativistic reflection model {\sc relxill\_nk}~\cite{r1,r2,r3}, which is designed to test possible deviations from the Kerr background~\cite{rmp}. {\sc relxill\_nk} does not include higher order disk images, and we can thus estimate the impact of the latter on the measurement of the model parameters. Our results suggest that the effect of higher order disk images is indeed very weak and difficult to estimate, so we confirm that the standard assumption of relativistic reflection models to ignore such radiation is perfectly acceptable for the quality of present and near future X-ray observations. Such a statement should be valid either when we want to measure the properties of the system assuming general relativity or when we want to test the Kerr metric around an accreting black hole.

The paper is organized as follows. In Section~\ref{s-hor}, we present the system that we have in mind and the origin of higher order disk images. In Section~\ref{s-line}, we calculate the impact of higher order disk images on a single iron line and on the full reflection spectrum. In Section~\ref{s-sim}, we simulate some observations with \textsl{Athena}~\cite{athena} and we fit the data with {\sc relxill\_nk} in order to estimate the impact of higher order disk images on the measurement of the properties of an accreting black hole, and in particular on tests of the Kerr hypothesis, when the reflection model employed in the fit does not include higher order disk images. Summary and conclusions are in Section~\ref{s-con}.

\begin{figure}[b]
\begin{center}
\includegraphics[type=pdf,ext=.pdf,read=.pdf,width=7.5cm]{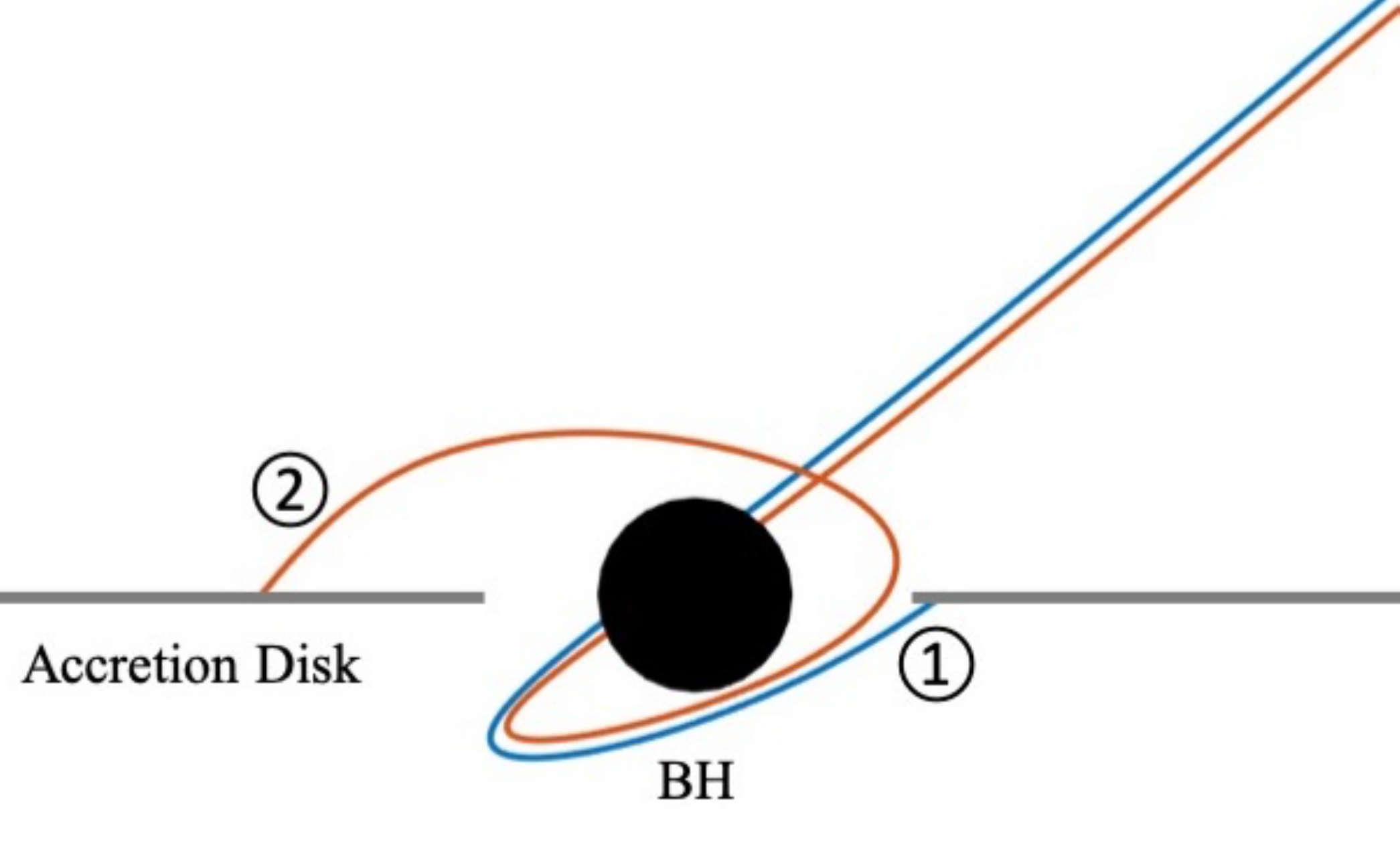}
\end{center}
\vspace{-0.3cm}
\caption{Black hole with accretion disk. The direct reflection spectrum is produced by reflection radiation that never crosses the equatorial plane between the black hole and the inner edge of the disk. The first order correction is generated on the surface of the other side of the disk and crosses the equatorial plane between the black hole and the inner edge of the disk once (1 in the figure). The second order correction crosses the equatorial plane between the black hole and the inner edge of the disk twice (2 in the figure). The $n$th order correction crosses the equatorial plane between the black hole and the inner edge of the disk $n$ times. \label{f-bh}}
\end{figure}


\section{Higher order disk images \label{s-hor}}

We consider a black hole surrounded by a geometrically thin and optically thick accretion disk, as shown in Fig.~\ref{f-bh}. The inner edge of the disk is assumed to be at the ISCO radius. The plunging region is the region between the inner edge of the disk and the black hole because the particles of the accreting gas reach the inner edge and then they are assumed to quickly plunge onto the black hole without appreciable emission of radiation. Depending on the properties of the accretion flow, in particular on the mass accretion rate, the plunging region can be either optically thin or optically thick. In the present paper, we consider the first case (optically thin) and we thus investigate the possible impact of radiation crossing the equatorial plane between the black hole and the inner edge of the accretion disk. If the plunging region is optically thick, the phenomenon of this paper is not present and we can instead have a reflection spectrum due to illumination of the plunging region by the corona. The impact of such an effect on X-ray reflection spectroscopy measurements will be presented in a forthcoming paper.

If the plunging region is optically thin, photons can travel around the black hole and cross the equatorial plane one or more times. The direct reflection spectrum is that generated by radiation that never crosses the plunging region and is the only one calculated in most relativistic reflection models. The first order correction is generated on the underside of the disk and crosses the plunging region once (trajectory~1 in Fig.~\ref{f-bh}). The second order correction is generated on the top side of the accretion disk and crosses the plunging region twice (trajectory~2 in Fig.~\ref{f-bh}). In general, we can talk about $n$th order correction, which is generated on the top side of the disk if $n$ is even and on the underside of the disk if $n$ is odd, and crosses the plunging region $n$ times. As we can expect, the contribution to the total spectrum quickly decreases as $n$ increases, and it gets more and more difficult to calculate it numerically with our code.


\section{Iron line shapes \label{s-line}}

The aim of this section is to determine the impact of higher order disk images. First, we simplify the reflection spectrum to a single iron line and we calculate the shape of the iron line, as detected by a distant observer, without and with higher order disk images. We assume that the iron line is at 6.4~keV in the rest-frame of the gas (neutral iron). The disk is assumed to be infinitesimally thin and perpendicular to the black hole spin. We employ the ray-tracing code used in Ref.~\cite{ss2}. The calculations of the spectrum of accretion disks with ray-tracing codes have been already discussed in literature (see, for instance, Refs.~\cite{book,c1,c2}), so here we only outline the main passages.

We consider an observer far from the black hole with viewing angle $i$, which is the angle between the line of sight of the observer and the black hole spin. We set a grid on the image plane of the observer and we fire a photon from every point of the grid to calculate its trajectory and where the photon hits the accretion disk. From the radial coordinate of the emission point on the accretion disk, we infer the gas velocity and, combined with the photon 4-momentum, we infer the redshift factor $g = E_{\rm obs}/E_{\rm e}$, where $E_{\rm obs}$ and $E_{\rm e}$ are, respectively, the photon energy measured by the distant observer and in the rest-frame of the gas. Integrating over the whole image of the accretion disk, we get the photon count as a function of $E_{\rm obs}$
\be\label{eq-n}
N(E_{\rm obs}) = \frac{1}{E_{\rm obs}} \int g^3 \, I_{\rm e} (E_e) \frac{dXdY}{D^2} \, ,
\ee  
where $I_{\rm e}$ is the specific intensity of the radiation at the emission point in the rest-frame of the gas, $X$ and $Y$ are the Cartesian coordinates in the observer plane, and $D$ is the distance between the black hole and the observer. For simplicity, we assume a power-law emissivity profile with emissivity index $q = 3$, namely $I_{\rm e} \propto r^{-3}$.

\begin{figure*}[h]
\begin{center}
\includegraphics[type=pdf,ext=.pdf,read=.pdf,width=17.5cm]{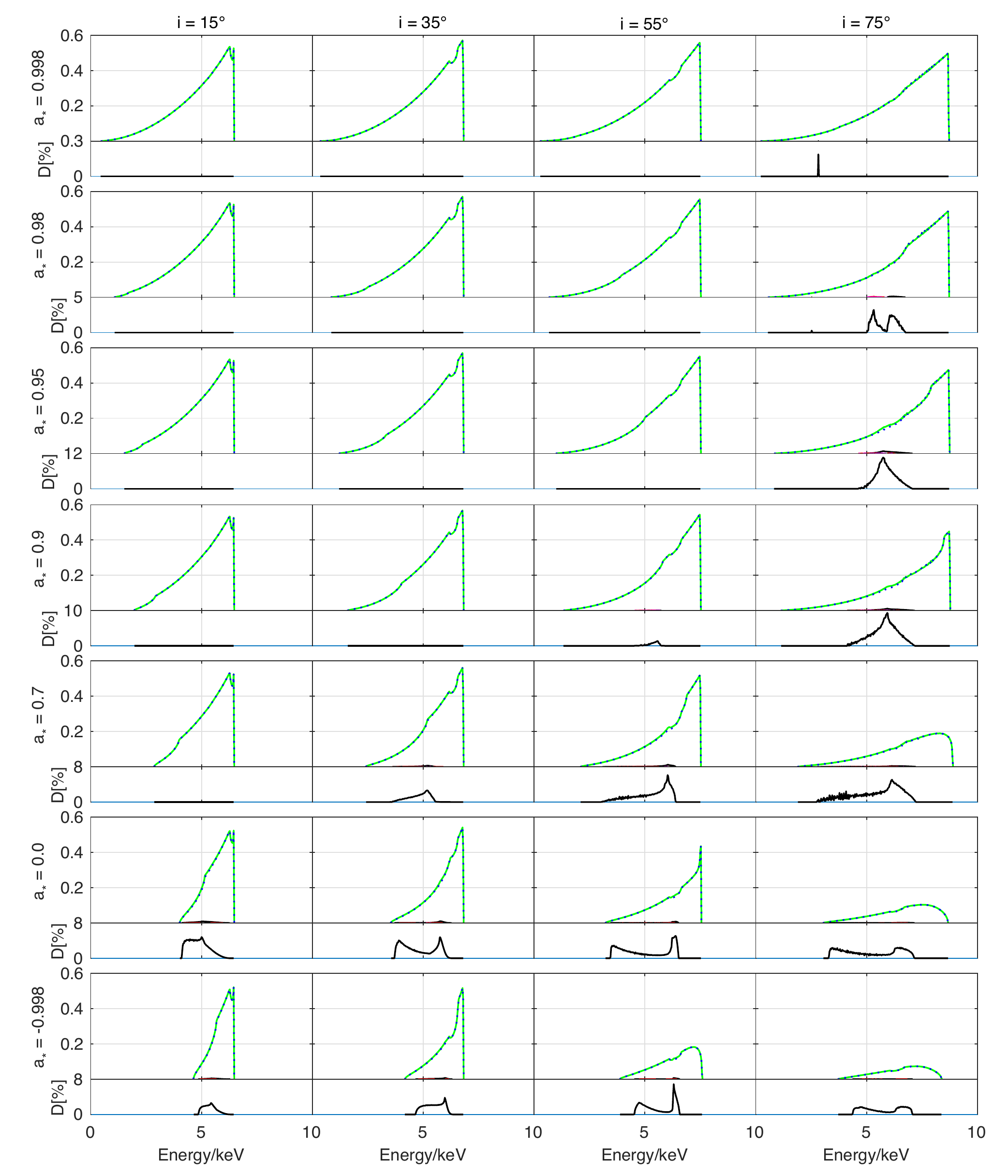}
\end{center}
\vspace{-0.5cm}
\caption{Iron lines shapes for different black hole spins and viewing angles. In every panel, the upper quadrant shows the total line (solid green curve), the line ignoring the radiation crossing the equatorial plane between the black hole and the inner edge of the disk (dotted blue curve), the contribution from the underside of the disk (solid black curve), the contribution from radiation circling the black hole once (dotted red curve), and the contribution from radiation circling the black hole two or more times (dashed-dotted magenta curve). The lower quadrant shows the relative difference between the solid green curve and the dotted blue one of the upper quadrant. \label{f-lines}}
\end{figure*}

The results of our calculations are shown in Fig.~\ref{f-lines} for seven different values of the spin parameter ($a_* = -0.998$, 0, 0.7, 0.9, 0.95, 0.98, 0.998) and four values of the viewing angle ($i = 15^\circ$, $35^\circ$, $55^\circ$, and $75^\circ$). Contributions from the direct image of the disk and from higher orders disk image are indicated with different colors in Fig.~\ref{f-lines}, but they are difficult to identify because the contribution is very small for higher order spectra. The solid green line indicates the total line (sum of all contributions). The dotted blue line indicates the contribution from the direct image of the disk, namely radiation that never crosses the equatorial plane. The solid black line indicates the radiation generated on the underside of the disk and crossing the plunging region once (first order contribution). The dotted red line (difficult to see) shows the contribution from the radiation crossing the plunging region twice. The dashed-dotted magenta line shows the contribution from higher order disk images, crossing the plunging region three or more times, but in most cases it cannot be seen. In every panel, the lower quadrant reports the relative difference between the total line and the dominant direct image contribution.

From Fig.~\ref{f-lines} it is easy to identify a trend in the contribution of higher order disk images: the contribution is negligible for high values of the spin parameters and low values of the viewing angle, while it is not negligible for low values of the spin parameter (which also includes negative values of $a_*$ corresponding to counterrotating disks) and/or high values of the viewing angle of the disk. However, for very high spin even in the case of high inclination angle the final impact of the reflection spectrum is very weak.

\begin{figure*}[t]
\begin{center}
\includegraphics[type=pdf,ext=.pdf,read=.pdf,width=17.0cm]{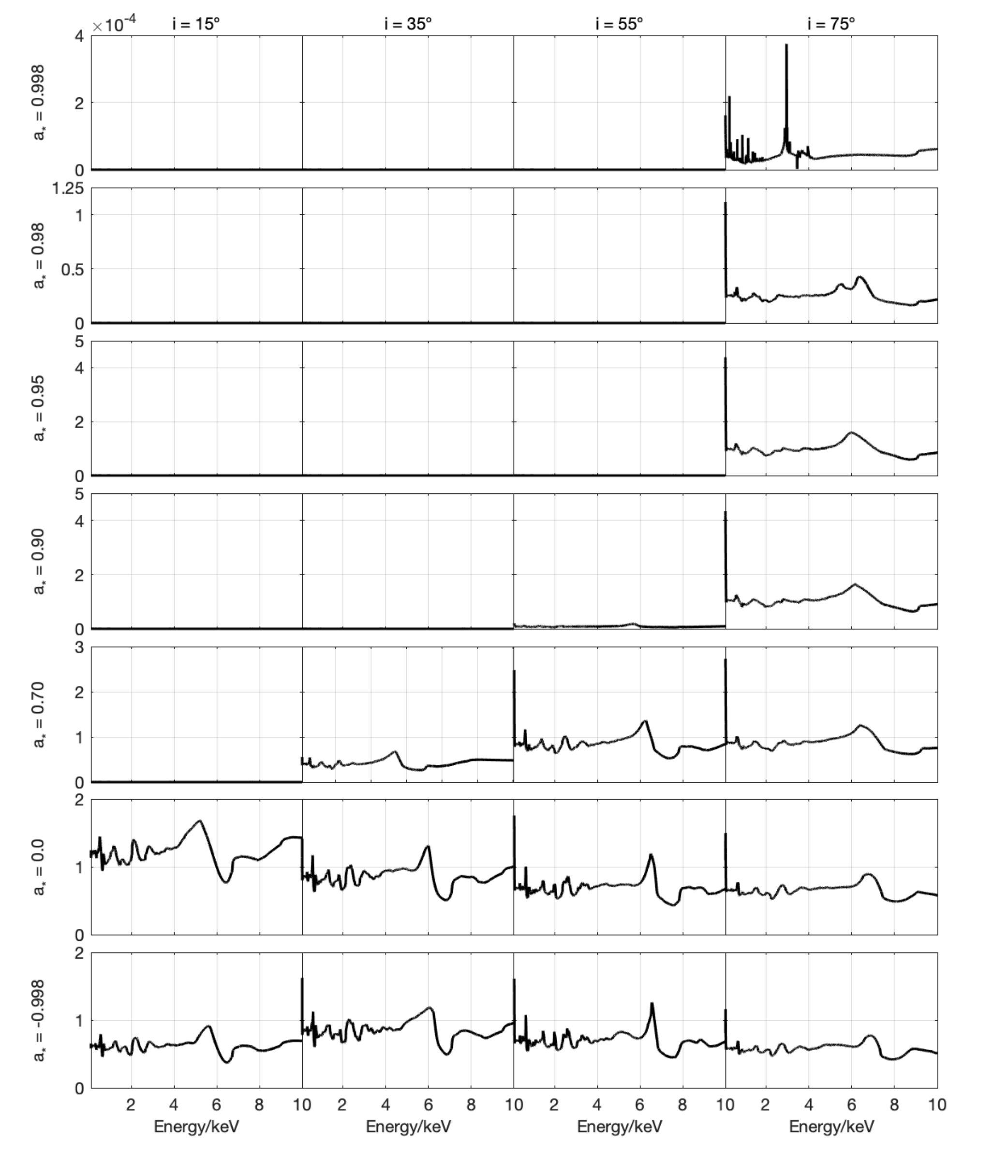}
\end{center}
\vspace{-0.9cm}
\caption{Relative difference (in \%) between reflection spectra with and without the radiation crossing the equatorial plane between the black hole and the inner edge of the disk for different black hole spins and viewing angles. \label{f-full}}
\end{figure*}

The calculations of iron line shapes without and with higher order disk images can be easily generalized to the full reflection spectrum. This is possible by changing $I_{\rm e}$ in Eq.~(\ref{eq-n}), replacing an iron line with a non-relativistic reflection spectrum calculated with {\sc xillver}~\cite{xill} and using the ray-tracing code as a convolution model to transform a spectrum at the emission point of the disk in the observed spectrum far from the source. Fig.~\ref{f-full} is the generalization of Fig.~\ref{f-lines} for the full reflection spectrum, but we only show the differences between spectra with and without higher order disk images because the full reflection spectrum has so many features that its visualization does not help the interpretation. The non-relativistic reflection spectrum is obtained with {\sc xillver} assuming that the incident radiation has photon index $\Gamma = 2.389$ and high energy cutoff $E_{\rm cut} = 300$~keV. We assume Solar iron abundance ($A_{\rm Fe} = 1$) and the ionization parameter is $\log\xi = 3.1$ ($\xi$ in units erg~cm~s$^{-1}$).

\section{Simulations \label{s-sim}}

In order to be more quantitative about the actual impact of higher order disk images on the measurements of the properties of a black hole, we simulate some observations with the XIFU instrument of \textsl{Athena}~\cite{athena}. If our results show that we cannot measure the difference between the reflection spectrum from the direct disk image and the full reflection spectrum that includes higher order contributions, we can conclude that the impact of higher order disk images is indeed negligible for observations with current and next generation X-ray missions.

We calculate the reflection spectrum with and without higher order contributions as done in the previous section. The reflection spectrum at the emission point is calculated with the non-relativistic reflection model {\sc xillver} and we adopt the same input parameters as before ($\Gamma = 2.389$, $E_{\rm cut} = 300$~keV, $\log\xi = 3.1$, $A_{\rm Fe} = 1$). The ray-tracing code is used as a convolution model. To maximize the difference between the spectra without (simulations~a) and with (simulations~b) higher order disk images, we consider the black hole spin parameters $a_* = 0.7$ (simulations~1) and 0.95 (simulations~2) and in both cases we set the viewing angle to 70$^\circ$. In the end, we have thus four spectra (1a, 1b, 2a, and 2b).

The simulated observations are obtained with the \verb|fakeit| command of XSPEC~\cite{xspec}, the response file and the background spectrum of \textsl{Athena}, and the four theoretical spectra discussed before. The total model is {\sc tbabs$\times$(powerlaw + reflection)}, where {\sc tbabs} takes the galactic absorption into account~\cite{tbabs}, {\sc powerlaw} describes the direct spectrum from the corona, and {\sc reflection} is our theoretical reflection spectrum (1a, 1b, 2a, or 2b). We assume the observation of a bright active galactic nucleus (AGN) and we set the photon flux to $1.4 \cdot 10^{-10}$~erg~cm$^{-2}$~s$^{-1}$. We consider an exposure time of 500~ks and we get a photon count of about 350~million photons in the 1-10~keV energy band.

\begin{figure*}[t]
\begin{center}
\includegraphics[type=pdf,ext=.pdf,read=.pdf,width=8.5cm]{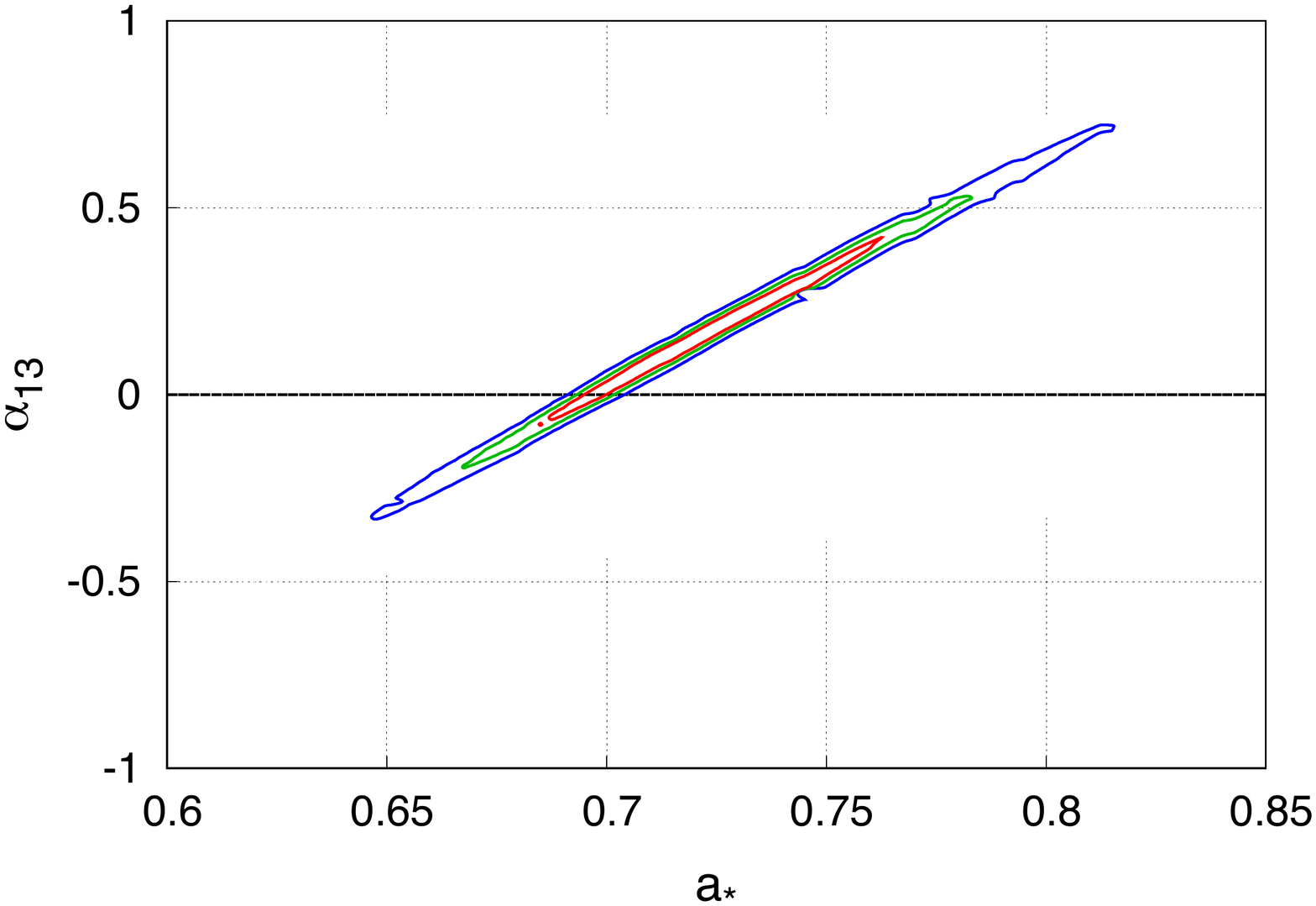}
\includegraphics[type=pdf,ext=.pdf,read=.pdf,width=8.5cm]{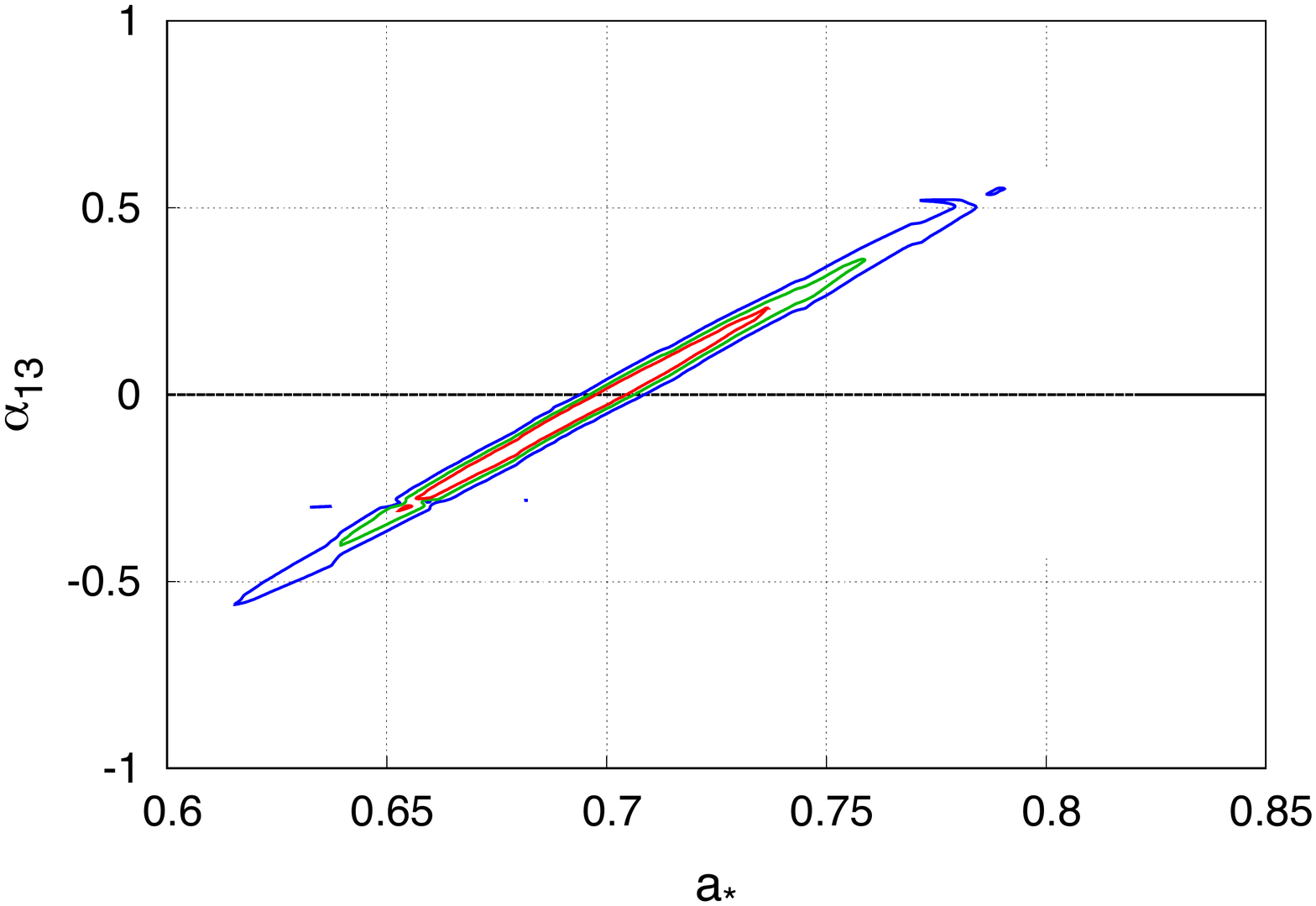}
\end{center}
\vspace{-1.2cm}
\caption{Constraints on the spin parameter $a_*$ and the deformation parameter $\alpha_{13}$ by fitting simulations 1a (left panel) and 1b (right panel) with the model {\sc tbabs$\times$relxill\_nk} that does not include the radiation crossing the equatorial plane between the inner edge of the disk and the black hole. \label{f-70}}
\vspace{0.0cm}
\begin{center}
\includegraphics[type=pdf,ext=.pdf,read=.pdf,width=8.5cm]{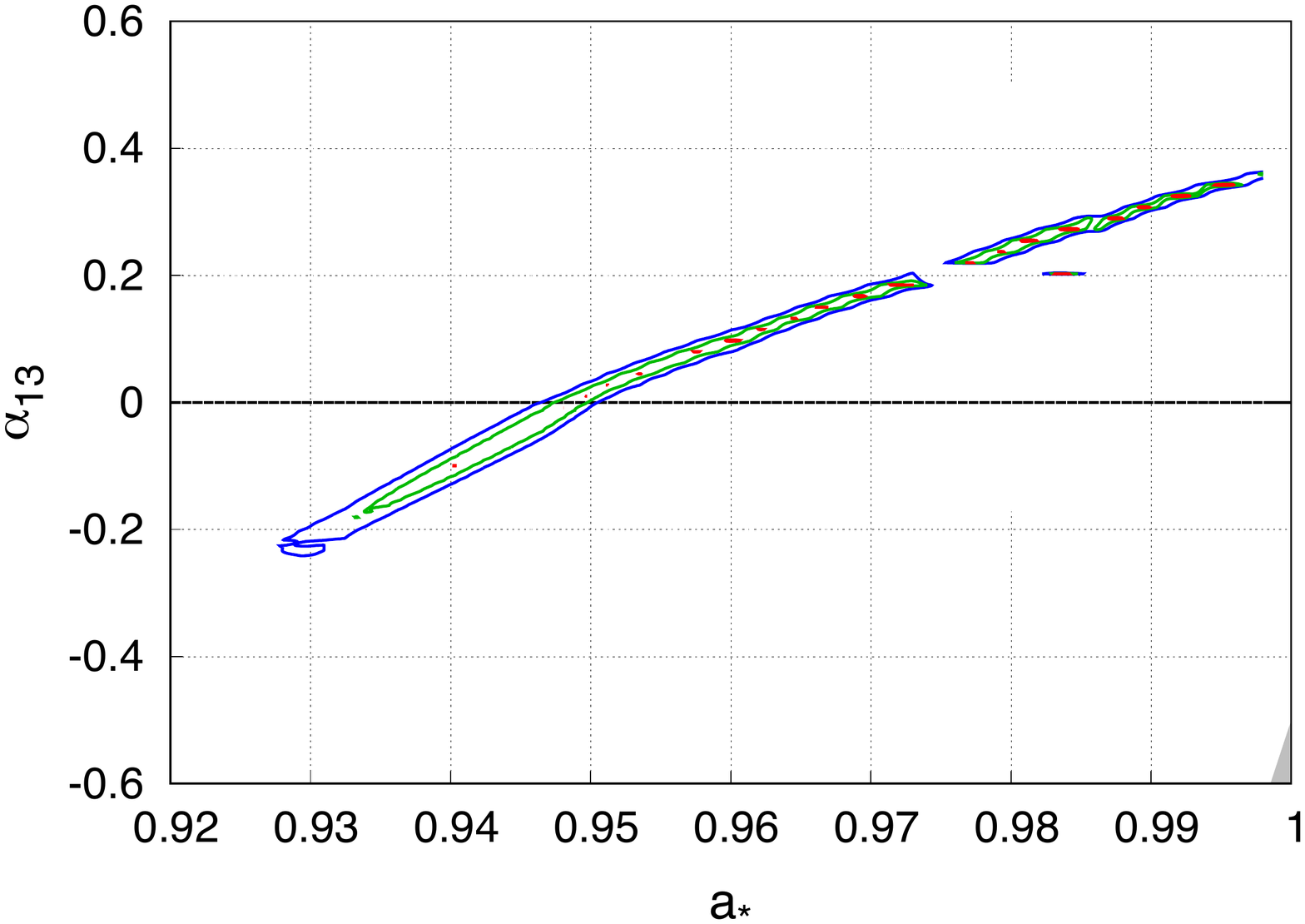}
\includegraphics[type=pdf,ext=.pdf,read=.pdf,width=8.5cm]{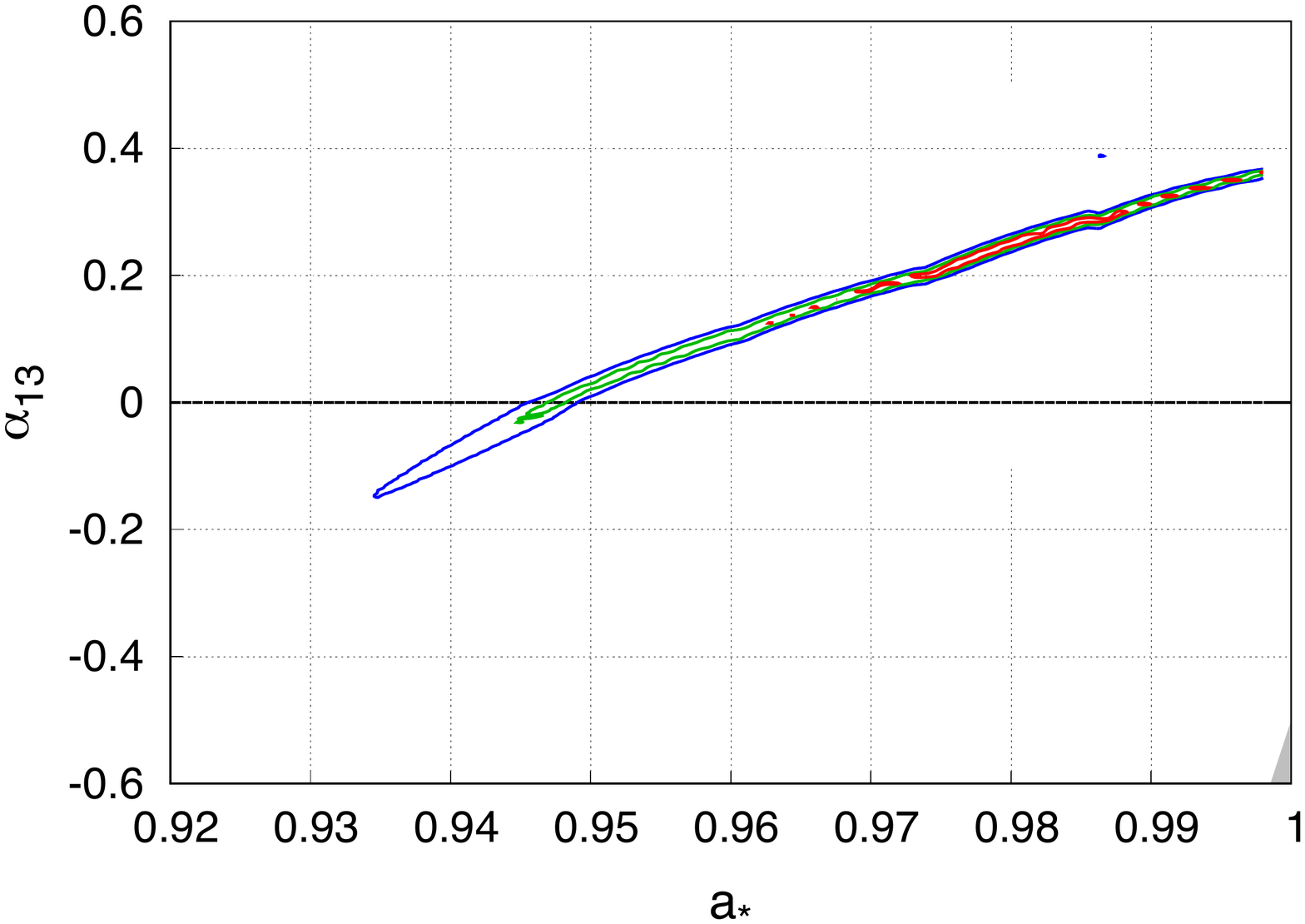}
\end{center}
\vspace{-1.2cm}
\caption{As in Fig.~\ref{f-70} for simulations 2a (left panel) and 2b (right panel). \label{f-95}}
\end{figure*}

The faked data are analyzed with the XSPEC model {\sc tbabs$\times$relxill\_nk}, where {\sc relxill\_nk} is our reflection model for a parametric black hole spacetime (see Appendix~\ref{s-app} for more details)~\cite{r1,r2,r3}. For the present study, we employ the Johannsen metric with the only non-vanishing deformation parameter $\alpha_{13}$, as it is the deformation parameter with the strongest impact on the reflection spectrum of the disk. The results of the fits for our four cases (1a, 1b, 2a, 2b) are shown in Tab.~\ref{t-fit}. In {\sc relxill\_nk} the reflection fraction $R_{\rm f}$ is free in order to describe both the reflection spectrum and the power-law spectrum from the corona. As we can see, in all simulations we recover the correct input parameters and there are not appreciable differences between the simulations in which higher disk images are or are not included. The measurement of the deformation parameter is consistent with zero (for simulation~2b, not at 90\% confidence level for one relevant parameter but at a slightly higher value). Since we are mainly interested in the systematic uncertainties related to tests of the Kerr metric, we also show in Fig.~\ref{f-70} and Fig.~\ref{f-95} the constraints on the spin and the deformation parameters after marginalizing over all other free parameters of the fit.

\begin{table*}
\centering
{\renewcommand{\arraystretch}{1.3}
\begin{tabular}{lcccccccccc}
\hline\hline
 & \multicolumn{2}{c}{Simulation~1a} && \multicolumn{2}{c}{Simulation~1b} & \multicolumn{2}{c}{Simulation~2a} && \multicolumn{2}{c}{Simulation~2b} \\
 & Input & Fit && Input & Fit & Input & Fit && Input & Fit \\
\hline
{\sc tbabs} &&&&&&&&& \\
$N_{\rm H} / 10^{20}$ cm$^{-2}$ & $6.74$ & $6.74^\star$ && $6.74$ & $6.74^\star$ & $6.74$ & $6.74^\star$ && $6.74$ & $6.74^\star$ \\
\hline
{\sc relxill\_nk} &&&&&&& \\
$q$ & $3$ & $3.003^{+0.008}_{-0.007}$ && $3$ & $2.970^{+0.007}_{-0.009}$ & $3$ & $3.019^{+0.014}_{-0.008}$ && $3$ & $2.980^{+0.013}_{-0.010}$ \\
$i$ [deg] & $70$ & $69.67^{+0.34}_{-0.16}$ && $70$ & $69.9^{+0.4}_{-0.4}$ & $70$ & $69.384^{+0.026}_{-0.018}$ && $70$ & $69.266^{+0.123}_{-0.015}$ \\
$a_*$ & $0.7$ & $0.72^{+0.03}_{-0.06}$ && $0.7$ & $0.69^{+0.05}_{-0.04}$ & $0.95$ & $0.99_{-0.03}^{\rm (P)}$ && $0.95$ & $0.998_{-0.024}^{\rm (P)}$ \\
$\alpha_{13}$ & $0$ & $0.2^{+0.3}_{-0.3}$ && $0$ & $0.0^{+0.3}_{-0.3}$ & $0$ & $0.32^{+0.05}_{-0.43}$ && $0$ & $0.35^{+0.02}_{-0.24}$ \\
$\log\xi$ & $3.1$ & $3.0956^{+0.0023}_{-0.0012}$ && $3.1$ & $3.0970^{+0.0022}_{-0.0021}$ & $3.1$ & $3.0945^{+0.0009}_{-0.0009}$ && $3.1$ & $3.0975^{+0.0012}_{-0.0006}$ \\
$A_{\rm Fe}$ & $1$ & $1.005^{+0.008}_{-0.006}$ && $1$ & $0.9995^{+0.0058}_{-0.0019}$ & $1$ & $1.007^{+0.006}_{-0.006}$ && $1$ & $0.9996^{+0.0041}_{-0.0015}$ \\
$\Gamma$ & $2.389$ & $2.3870^{+0.0007}_{-0.0007}$ && $2.389$ & $2.3885^{+0.0005}_{-0.0008}$ & $2.389$ & $2.3882^{+0.0003}_{-0.0003}$ && $2.389$ & $2.3887^{+0.0005}_{-0.0006}$ \\
$E_{\rm cut}$ [keV] & $300$ & $300^\star$ && $300$ & $300^\star$ & $300$ & $300^\star$ && $300$ & $300^\star$ \\
$R_{\rm f}$ & -- & $0.445^{+0.006}_{-0.003}$ && -- & $0.445^{+0.009}_{-0.005}$ & -- & $0.4426^{+0.0034}_{-0.0022}$ && -- & $0.447^{+0.004}_{-0.003}$ \\
\hline
Higher order & No & -- && Yes & -- & No & -- && Yes & -- \\
\hline
$\chi^2/\nu$ && $\quad 24796.77/24510 \quad$ &&& $\quad 25041.82/24515 \quad$ && $\quad 24539.83/24494 \quad$ &&& $\quad 24610.57/24509 \quad$ \\
&& = 1.01170 &&& = 1.02149 && = 1.00187 &&& = 1.00414 \\
\hline\hline
\end{tabular}}
\caption{Input parameters and best-fit values for simulations~1a, 1b, 2a, and 2b. The reported uncertainties correspond to 90\% confidence level for one relevant parameter. $^\star$ indicates that the parameter is frozen in the fit. In the simulations, we do not include (do include) the radiation crossing the equatorial plane between the inner edge of the disk and the black hole when ``Higher order'' is ``No'' (``Yes'').} \label{t-fit}
\end{table*}

\section{Discussion and conclusions \label{s-con}}

If the region between the inner edge of the accretion disk and the black hole is optically thin, the total reflection spectrum of the source should be the sum of the direct reflection spectrum (radiation that never crosses the equatorial plane) and higher order disk images. The latter are produced by the other side of the disk and by radiation circling the black hole one or more times. The phenomenon is due to the strong light-bending in the vicinity of the compact object and is usually neglected in current relativistic reflection models. In this paper we have investigated the impact of this radiation on the reflection spectrum and, in particular, on the measurements of the properties of the accreting black hole.

The impact of this radiation on the shape of an iron line and on the full reflection spectrum is summarized, respectively, in Fig.~\ref{f-lines} and Fig.~\ref{f-full}, where we show the results for different black hole spins (we assume that the inner edge is at the ISCO radius) and different viewing angles of the disk. For a Schwarzschild black hole ($a_* = 0$), the difference between an iron line without and with higher order disk images is at the level of 3-6\%, while the difference is at the level of 1-2\% for the full reflection spectrum. This is not very sensitive to the exact viewing angle of the disk, but the trend is that the discrepancy is larger for lower viewing angles and smaller for higher viewing angles. In the case of a counterrotating disk, we find similar results. If, on the contrary, we consider corotating disks and we increase the value of the spin parameter, so the size of the plunging region decreases because the inner edge of the disk gets closer to the black hole, we see that the discrepancy between without and with higher order disk images decreases for low inclination angles and increases for high inclination angles. For $a_* = 0.9$, it seems that the discrepancy is negligible except for inclination angles like 70$^\circ$ or higher. For very high values of the black hole spin parameters (like $a_* = 0.98$ and 0.998), the difference between without and with higher order disk images becomes very small even for very large viewing angles.

For a more quantitative evaluation of the actual impact of higher order disk images on current and near future measurements, we have simulated some observations with \textsl{Athena}. More specifically, we have considered four cases, called, respectively, 1a, 1b, 2a, and 2b. For the cases 1 (2), the spin parameter was set to 0.7 (0.95), because for very high spins the effect of higher order radiation is surely weaker. For the cases a (b), we did not include (did include) higher order disk images in the simulated spectrum. We considered observation of a bright AGN for 500~ks (around 350~million photons in the 1-10~keV range) and then we fit the faked data with {\sc relxill\_nk}. Our fit can recover the correct input values and we do not see any appreciable difference between the simulations in which we included and did not include higher order disk images. This is the case also for the deformation parameter $\alpha_{13}$, as shown in Fig.~\ref{f-70} and Fig.~\ref{f-95}, respectively for $a_* = 0.7$ and $a_* = 0.95$. Note that the input viewing angle of the disk is $i = 70^\circ$, which is definitively too high for an AGN showing a strong reflection spectrum. For such high values of the viewing angle, the inner part of the accretion disk is usually obscured and we do not see the reflection spectrum. Considering that typical values are $i < 40^\circ$, the impact of higher order reflection for sources used to test general relativity using X-ray reflection spectroscopy should be even weaker and thus completely negligible for observations with current and next generation X-ray missions.

As pointed out in the introduction section, we can expect that other modeling simplifications can introduce larger uncertainties in the estimate of the properties of the system. One may thus question whether these other simplifications (like, for instance, the thickness of the disk or the coronal geometry) can significantly change our conclusions on the impact of higher order disk images when they are properly taken onto account. We do not expect changes in our conclusions. The calculation of higher order disk images indeed relies on ray-tracing calculations (in which there are no simplifications) and on the spectrum from the disk. For the latter, we may have a discrepancy between synthetic and actual spectra, but the uncertainties are at most of order unity, so the contribution from higher order disk images can somewhat change but cannot increase by an order of magnitude with respect to our estimates. It is small and it should remain small even when the other systematic uncertainties are taken into account.


\vspace{0.5cm}

{\bf Acknowledgments --}
This work was supported by the Innovation Program of the Shanghai Municipal Education Commission, Grant No.~2019-01-07-00-07-E00035, the National Natural Science Foundation of China (NSFC), Grant No.~11973019, and Fudan University, Grant No.~IDH1512060.


\appendix

\section{Johannsen metric \label{s-app}}

{\sc relxill\_nk} is an extension of the {\sc relxill} package~\cite{rx1,rx2} to parametric black hole spacetimes. In this model, the background metric is obtained by deforming the Kerr solution. The new metric is specified by the mass $M$ and the spin angular momentum $J$ of the compact object as well as by a number of ``deformation parameters'', which are introduced to quantify deviations from the Kerr geometry. The Kerr solution is exactly recovered when all deformation parameters vanish. Within the spirit of a null experiment, we fit the observational data with this model and we try to constrain the value of the deformation parameters. The Kerr hypothesis is verified if the data require vanishing deformation parameters.

In the present paper, we have employed the Johannsen metric~\cite{tj}. In Boyer-Lindquist-like coordinates, the line element reads
\be\label{eq-jm}
ds^2 &=&-\frac{\tilde{\Sigma}\left(\Delta-a^2A_2^2\sin^2\theta\right)}{B^2}dt^2
+\frac{\tilde{\Sigma}}{\Delta}dr^2+\tilde{\Sigma} d\theta^2 \nonumber\\
&&-\frac{2a\left[\left(r^2+a^2\right)A_1A_2-\Delta\right]\tilde{\Sigma}\sin^2\theta}{B^2}dtd\phi \nonumber\\
&&+\frac{\left[\left(r^2+a^2\right)^2A_1^2-a^2\Delta\sin^2\theta\right]\tilde{\Sigma}\sin^2\theta}{B^2}d\phi^2 \, , 
\qquad
\ee
where $M$ is the black hole mass, $a = J/M$, $J$ is the black hole spin angular momentum, $\tilde{\Sigma} = \Sigma = f$, and
\be
\Sigma &=& r^2 + a^2 \cos^2\theta \, , \nonumber\\
\Delta &=& r^2 - 2 M r + a^2 \, , \nonumber\\
B &=& \left(r^2+a^2\right)A_1-a^2A_2\sin^2\theta \, .
\ee
The functions $f$, $A_1$, $A_2$, and $A_5$ are defined as
\be
f &=& \sum^\infty_{n=3} \epsilon_n \frac{M^n}{r^{n-2}} \, , \nonumber\\
A_1 &=& 1 + \sum^\infty_{n=3} \alpha_{1n} \left(\frac{M}{r}\right)^n \, , \nonumber\\
A_2 &=& 1 + \sum^\infty_{n=2} \alpha_{2n}\left(\frac{M}{r}\right)^n \, , \nonumber\\
A_5 &=& 1 + \sum^\infty_{n=2} \alpha_{5n}\left(\frac{M}{r}\right)^n \, .
\ee
$\{ \epsilon_n \}$, $\{ \alpha_{1n} \}$, $\{ \alpha_{2n} \}$, and $\{ \alpha_{5n} \}$ are four infinite sets of deformation parameters without constraints from the Newtonian limit and Solar System experiments. In the study presented in this paper, we have only considered the deformation parameter $\alpha_{13}$, because it has the strongest impact, among all the deformation parameters, on the reflection spectrum~\cite{r1}. In order to avoid spacetimes with pathological properties (naked singularities, closed time-like curves, etc.), we require that $| a_* | \le 1$ (as in the standard Kerr case) and the following constraint on $\alpha_{13}$ (see Ref.~\cite{564} for its derivation)
\be
\label{eq-constraints}
\alpha_{13} > - \frac{1}{2} \left( 1 + \sqrt{1 - a^2_*} \right)^4 \, .
\ee



\begin{thebibliography}{99}

\bibitem{o1} 
  A.~C.~Fabian, M.~J.~Rees, L.~Stella and N.~E.~White,
  Mon.\ Not.\ Roy.\ Astron.\ Soc.\  {\bf 238}, 729 (1989).
  
\bibitem{o2} 
  Y.~Tanaka {\it et al.},
  Nature {\bf 375}, 659 (1995).
  
\bibitem{o3} 
  K.~Nandra, I.~M.~George, R.~F.~Mushotzky, T.~J.~Turner and T.~Yaqoob,
  Astrophys.\ J.\  {\bf 477}, 602 (1997)
  [astro-ph/9606169].    

\bibitem{o4} 
  J.~M.~Miller,
  Ann.\ Rev.\ Astron.\ Astrophys.\  {\bf 45}, 441 (2007)
  [arXiv:0705.0540 [astro-ph]].

\bibitem{o5} 
  K.~Nandra, P.~M.~O'Neill, I.~M.~George and J.~N.~Reeves,
  Mon.\ Not.\ Roy.\ Astron.\ Soc.\  {\bf 382}, 194 (2007)
  doi:10.1111/j.1365-2966.2007.12331.x
  [arXiv:0708.1305 [astro-ph]].  
  
\bibitem{rev} 
  C.~Bambi,
  Annalen Phys.\  {\bf 530}, 1700430 (2018)
  [arXiv:1711.10256 [gr-qc]].
  
\bibitem{book} 
  C.~Bambi,
  {\it Black Holes: A Laboratory for Testing Strong Gravity} (Springer Singapore, 2017),
  doi:10.1007/978-981-10-4524-0  
  
\bibitem{s1} 
  L.~W.~Brenneman and C.~S.~Reynolds,
  Astrophys.\ J.\  {\bf 652}, 1028 (2006)
  [astro-ph/0608502].  
  
\bibitem{s1b} 
  D.~J.~Walton, E.~Nardini, A.~C.~Fabian, L.~C.~Gallo and R.~C.~Reis,
  Mon.\ Not.\ Roy.\ Astron.\ Soc.\  {\bf 428}, 2901 (2013)
  [arXiv:1210.4593 [astro-ph.HE]].  
  
\bibitem{s2} 
  T.~Dauser, J.~Garcia, J.~Wilms, M.~Bock, L.~W.~Brenneman, M.~Falanga, K.~Fukumura and C.~S.~Reynolds,
  Mon.\ Not.\ Roy.\ Astron.\ Soc.\  {\bf 430}, 1694 (2013)
  [arXiv:1301.4922 [astro-ph.HE]].  
  
\bibitem{s3} 
  C.~S.~Reynolds,
  Space Sci.\ Rev.\  {\bf 183}, 277 (2014)
  [arXiv:1302.3260 [astro-ph.HE]].

\bibitem{s4} 
  L.~Brenneman,
  {\it Measuring Supermassive Black Hole Spins in Active Galactic Nuclei} (Springer, 2013),
  doi:10.1007/978-1-4614-7771-6
  arXiv:1309.6334 [astro-ph.HE].  
  
\bibitem{t1} 
  Z.~Cao, S.~Nampalliwar, C.~Bambi, T.~Dauser and J.~A.~Garcia,
  Phys.\ Rev.\ Lett.\  {\bf 120}, 051101 (2018)
  [arXiv:1709.00219 [gr-qc]].
    
\bibitem{t1b} 
  Y.~Xu, S.~Nampalliwar, A.~B.~Abdikamalov, D.~Ayzenberg, C.~Bambi, T.~Dauser, J.~A.~Garcia and J.~Jiang,
  Astrophys.\ J.\  {\bf 865}, 134 (2018)
  [arXiv:1807.10243 [gr-qc]].    
    
\bibitem{t2} 
  A.~Tripathi, S.~Nampalliwar, A.~B.~Abdikamalov, D.~Ayzenberg, C.~Bambi, T.~Dauser, J.~A.~Garcia and A.~Marinucci,
  Astrophys.\ J.\  {\bf 875}, 56 (2019)
  [arXiv:1811.08148 [gr-qc]]. 
  
\bibitem{t3} 
  A.~Tripathi {\it et al.},
  Astrophys.\ J.\  {\bf 874}, 135 (2019)
  [arXiv:1901.03064 [gr-qc]].   
  
\bibitem{t4} 
  Y.~Zhang, A.~B.~Abdikamalov, D.~Ayzenberg, C.~Bambi and S.~Nampalliwar,
  Astrophys.\ J.\  {\bf 884}, 147 (2019)
  [arXiv:1907.03084 [gr-qc]].  
  
\bibitem{ss1} 
  H.~Liu, A.~B.~Abdikamalov, D.~Ayzenberg, C.~Bambi, T.~Dauser, J.~A.~Garcia and S.~Nampalliwar,
  Phys.\ Rev.\ D {\bf 99}, 123007 (2019)
  [arXiv:1904.08027 [gr-qc]].  
  
\bibitem{ss2}
  S.~Riaz, D.~Ayzenberg, C.~Bambi and S.~Nampalliwar,
  Mon.\ Not.\ Roy.\ Astron.\ Soc.\  {\bf 491}, 417 (2020)
  [arXiv:1908.04969 [astro-ph.HE]].  
  
\bibitem{ss3} 
  B.~Zhou, A.~Tripathi, A.~B.~Abdikamalov, D.~Ayzenberg, C.~Bambi, S.~Nampalliwar and M.~Zhou,
  arXiv:1908.05177 [gr-qc].  
  
\bibitem{ss4} 
  S.~Riaz, D.~Ayzenberg, C.~Bambi and S.~Nampalliwar,
  arXiv:1911.06605 [astro-ph.HE].  
  
\bibitem{andrzej1} 
  A.~Niedzwiecki, A.~A.~Zdziarski and M.~Szanecki,
  Astrophys.\ J.\  {\bf 821}, L1 (2016)
  [arXiv:1602.09075 [astro-ph.HE]].

\bibitem{andrzej2} 
  A.~Niedzwiecki and A.~A.~Zdziarski,
  Mon.\ Not.\ Roy.\ Astron.\ Soc.\  {\bf 477}, 4269 (2018)
  [arXiv:1803.03781 [astro-ph.HE]].  
  
\bibitem{r1} 
  C.~Bambi, A.~Cardenas-Avendano, T.~Dauser, J.~A.~Garcia and S.~Nampalliwar,
  Astrophys.\ J.\  {\bf 842}, 76 (2017)
  [arXiv:1607.00596 [gr-qc]].  
  
\bibitem{r2} 
  A.~B.~Abdikamalov, D.~Ayzenberg, C.~Bambi, T.~Dauser, J.~A.~Garcia and S.~Nampalliwar,
  Astrophys.\ J.\  {\bf 878}, 91 (2019)
  [arXiv:1902.09665 [gr-qc]].  

\bibitem{r3} 
  A.~B.~Abdikamalov, D.~Ayzenberg, C.~Bambi and S.~Nampalliwar,
  MDPI Proc.\  {\bf 17}, 7 (2019)
  [arXiv:1908.10152 [gr-qc]].
  
\bibitem{rmp} 
  C.~Bambi,
  Rev.\ Mod.\ Phys.\  {\bf 89}, 025001 (2017)
  [arXiv:1509.03884 [gr-qc]].  
  
\bibitem{athena} 
  K.~Nandra {\it et al.},
  arXiv:1306.2307 [astro-ph.HE].  

\bibitem{c1} 
  C.~Bambi,
  Astrophys.\ J.\  {\bf 761}, 174 (2012)
  [arXiv:1210.5679 [gr-qc]].

\bibitem{c2} 
  C.~Bambi,
  Phys.\ Rev.\ D {\bf 87}, 023007 (2013)
  [arXiv:1211.2513 [gr-qc]].
  
\bibitem{xill} 
  J.~Garcia and T.~Kallman,
  Astrophys.\ J.\  {\bf 718}, 695 (2010)
  [arXiv:1006.0485 [astro-ph.HE]].
  
\bibitem{xspec} 
  K.~A.~Arnaud,
  Astronomical Data Analysis Software and Systems V, {\bf 101}, 17 (1996).  

\bibitem{tbabs} 
  J.~Wilms, A.~Allen and R.~McCray,
  Astrophys.\ J.\  {\bf 542}, 914 (2000)
  [astro-ph/0008425].

\bibitem{rx1} 
  J.~Garcia, T.~Dauser, C.~S.~Reynolds, T.~R.~Kallman, J.~E.~McClintock, J.~Wilms and W.~Eikmann,
  Astrophys.\ J.\  {\bf 768}, 146 (2013)
  [arXiv:1303.2112 [astro-ph.HE]].  
  
\bibitem{rx2} 
  J.~García {\it et al.},
  Astrophys.\ J.\  {\bf 782}, 76 (2014)
  [arXiv:1312.3231 [astro-ph.HE]].  
  
\bibitem{tj} 
  T.~Johannsen,
  Phys.\ Rev.\ D {\bf 88}, 044002 (2013)
  [arXiv:1501.02809 [gr-qc]].  
  
\bibitem{564} 
  A.~Tripathi, S.~Nampalliwar, A.~B.~Abdikamalov, D.~Ayzenberg, J.~Jiang and C.~Bambi,
  Phys.\ Rev.\ D {\bf 98}, 023018 (2018)
  [arXiv:1804.10380 [gr-qc]].  

\end{thebibliography}
\end{document}